\documentclass[twocolumn,twoside,slac_two, nofootinbib]{revtex4-1}
\usepackage{graphicx}
\usepackage{fancyhdr}
\usepackage{amsmath}
 \usepackage{amssymb}
 \usepackage{amsfonts}
 \usepackage{eucal}

\newcommand{\be}{\begin{equation}}
\newcommand{\bse}{\begin{subequations}}
\newcommand{\ese}{\end{subequations}}
\newcommand{\bea}{\begin{eqnarray}}
\newcommand{\eea}{\end{eqnarray}}
\newcommand{\ba}{\begin{array}}
\newcommand{\ea}{\end{array}}
\newcommand{\ee}{\end{equation}}

\def\Pl{{\rm Pl}}
\def\hp{h_+}
\def\hc{h_{\times}}
\def\half{\frac{1}{2}}

\def\yboxit#1#2{\vbox{\hrule height #1 \hbox{\vrule width #1
    \vbox{#2}\vrule width #1 }\hrule height #1 }}
\def\fillbox#1{\hbox to #1{\vbox to #1{\vfil}\hfil}}
\def\ybox{\yboxit{0.4pt}{\fillbox{8pt}}\hskip-0.4pt}
\def\VEV#1{\langle{ #1} \rangle}

\begin{document}


\title{\centerline{Gravi-Leptogenesis:}
\centerline{Leptogenesis from Gravity Waves in Pseudo-scalar Driven Inflation Models}}
%

\author{S. Alexander}
\affiliation{Physics Department and the Institute for Gravitational Physics and geometry,\\
The Pennsylvania State University,
104 Davey Lab, University Park, PA 16802, U.S.A }
\author{M. Peskin}
\affiliation{SLAC, Stanford University,
2575 Sand Hill Road, Menlo Park, CA 94025, U.S.A}
\author{M. M. Sheikh-Jabbari}
\affiliation{Institute for studies in theoretical Physics and Mathematics (IPM),\\
P.O.Box 19395-5531, Tehran, IRAN}

\begin{abstract}
In this talk we present a mechanism for leptogenesis  which is based on gravity waves
produced during inflation. We show that when inflation is driven by a pseudo-scalar field
the metric perturbations generated during inflation can become birefringent, therefore
 giving a non-vanishing contribution to the gravitational triangle anomaly
and sourcing lepton anti-lepton asymmetry.  As this asymmetry is sourced by
the fields which are active during inflation, it is not washed out or diluted by inflation.
The amount of matter asymmetry generated in our model can be of realistic size for
the parameters within the range of some inflationary scenarios and grand
unified theories. This talk
is based on  \cite{APS} which has appeared on the arXiv as hep-th/0403069.

\end{abstract}
\maketitle

\thispagestyle{fancy}

\section{Introduction}

One of the puzzles of astroparticle physics and cosmology, which has been around for more
than half a century and since the existence of antimatter and anti-particles
established in collider experiments,
has been to explain if our universe is mostly made out of matter, why
and how this has happened during the course of the evolution of the universe starting from a
symmetric soup of matter and antimatter soon after the Big Bang.

The fact that in the parts of the
universe visible to us there is an excess
of matter over antimatter has now been backed by
the recent determinations of the cosmological parameters
from the cosmic microwave background observations and  the WMAP experiment.
Quantitatively this asymmetry is usually given through the  ratio
of excess of baryon density to the photon density \cite{WMAP}
\be\label{baryond}
\frac{n_B}{n_\gamma} = (6.5\pm 0.4)\times 10^{-10}\ ,
\ee
where $n_B= n_b- n_{\bar b}$ and $n_\gamma$ is the number density of photons.
This ratio has the nice property that is time independent, as
the evolution of the $n_b$ and $n_\gamma$ with the cosmic Hubble expansion are the same.
This is a small number, but at the same time it is large enough to be a
puzzle for models of particle physics.

About forty years ago Sakharov stated the three necessary conditions to generate a
matter-antimatter asymmetry dynamically from a symmetric initial
conditions~\cite{Sakharov}:\\
{\it i)} our particle physics model should have baryon number violating
vertices.\\
{\it ii)} CP should be violated.\\
{\it iii)} CP and baryon number violating interactions should be active
at a time when the universe is out of thermal equilibrium.

A baryon excess this large cannot be
produced in the early universe within the Standard Model (SM) of particle physics
\cite{HSather}. This is due to the fact that baryon number violating interactions in
the Standard Model are loop suppressed and the only source of CP violation in the hadronic
sector is in the Dirac phase of the CKM mixing matrix, which is not enough for explaining
the baryon asymmetry observed today. Moreover, assuming that the
scale of inflation is larger than TeV, {\it i.e.} the SM is at work after inflation is ended,
the out-of-equilibrium condition can be created at phase transitions or through
 late decay of massive particles.  The most attractive
 choice for a phase transition
 is that associated with electroweak symmetry breaking.  However, that
 phase transition is probably not sufficiently strongly first-order.

Although what is observed is a baryon asymmetry, since the 1980's
it has been realized that the standard weak interactions contain processes,
mediated by {\it sphalerons} ($SU(2)$ instantons),
which interconvert baryons and leptons and are
thermally activated at temperatures greater than 1 TeV.  Thus, we can also
create the baryon asymmetry by creating net lepton number at high temperature
through out-of-equilibrium and CP-asymmetric processes \cite{KRS,FY}.  Scenarios of this
type are known as {\it leptogenesis}.

To use the possibility of lepton asymmetry  and sphalerons we, however, need to fulfill
Sakharov conditions for leptons. Within the usual SM this does not solve the issue and
we are hence forced to associate the observed baryon asymmetry of the universe to physics
beyond the SM.

Most of the allowed parameter space of the
minimal supersymmetric Standard Model has already been excluded and
large CP violating phases are strongly
constrained in supersymmetric models \cite{SUSYphases} though they still could appear
 in the neutrino Yukawa couplings that are used in the  Fukugita-Yanagida
scenario for
leptogenesis \cite{FY}.
Models that can explain the baryon excess typically
involve exotic nonstandard physics, CP violating couplings in the Higgs
or supersymmetry sectors or in the couplings of the heavy
neutral leptons associated with neutrino mass \cite{troddenreview}.
 In any event, there is good reason to seek more effective
sources of CP-violating out-of-equilibrium physics.

Here we present a new mechanism for the creation of the matter-antimatter
asymmetry, one associated with gravitational fluctuations created during
cosmological inflation \cite{APS}.

\section{Outline of the mechanism}

Let us first spell out how the three Sakharov conditions are realized in our model of
 matter-antimatter asymmetry, {\it gravi-leptogenesis}.

\subsection{Lepton number violation}

The lepton number violation in our model comes from triangle anomaly.
As it is well-known \cite{AGW},
the lepton number current, and
hence the total fermion number current, has a gravitational anomaly in the
Standard Model.  Explicitly,
\be\label{Jlepton}
     \partial_\mu J^\mu_\ell  =  \frac{N}{16\pi^2}   R  \tilde R
\ee
where
\begin{subequations}
\begin{align}
J^\mu_\ell &=   \sum_{i=L,R} \bar \ell_i\gamma^\mu \ell_i + \bar \nu_i \gamma^\mu \nu_i \
,\\
R\tilde R & = \half \epsilon^{\alpha\beta\gamma\delta}
 R_{\alpha\beta \rho\sigma} R_{\gamma\delta}{}^{\rho\sigma} \ ,
\end{align}
\end{subequations}
and $N=N_L-N_R$, which is three in the Standard Model.
The anomaly is a consequence of an imbalance between left- and right-handed leptons.
In general when heavy right-handed neutrinos are also added to the Standard Model, as is done
in the seesaw mechanism for explaining the smallness of the neutrino mass,
\eqref{Jlepton} will be correct in an effective theory valid below a scale $\mu$,
of order of the right-handed neutrino mass.
More concretely $N$ can in general be a function of energy. At low energies, below the
right-handed neutrino mass scale $N=3$. At higher energies, $N$ could be
anywhere between zero to three, depending on the details of the particle physics model
invoked. In the usual seesaw scenarios with three right handed neutrinos,
$\mu$ can be as large as $10^{14}$, for energies below $\mu$ $N=3$, and above that $N=0$
\cite{seesaw1,seesaw2}. In any case, here we do not restrict ourselves to a specific
particle physics model, and keep an open mind on larger values of $\mu$.

\subsection{CP violation}

The need for CP violation manifests itself in our model through the fact that a
non-zero lepton number generation can be achieved when
 $\langle R\tilde R\rangle$ is non-vanishing. As we will show explicitly,
  $R\tilde R$ receives a contribution with a definite sign
from gravitational fluctuations produced during
 inflation, which is driven by a {\it pseudo-scalar field}.
In other words, CP-violation in our model arises from the inflaton field with a
CP-odd component. The simplest model of this kind is when we have a single field
inflation and a pseudo-scalar $\phi$ as the inflaton, known as natural inflation, however it
can be incorporated to have multiple axions such as in N-inflation models \cite{Shamit}.   In fact, these models of inflation fit quite nicely into extensions of
the standard model and in string inspired inflation.  Note that during inflation
the expectation value of the inflaton is non-vanishing and is rolling in time.

Such models of inflation can be naturally achieved if the inflaton is a
complex modulus field such as one finds in supergravity or superstring models.
In order to use these models, however, we need to make sure that they have flat enough
potentials required for (slow-roll) inflation.

The imaginary part $\phi$ of this field (which we henceforth call an `axion') can couple
to gravity through an interaction
\be\label{axioncoupling}
     \Delta {\cal L} =    F(\phi)  R  \tilde R  \ ,
\ee
where $F$ is odd in $\phi$. Under $P$ and $CP$
\[
\phi \to -\phi \ \& \ \ F(\phi)\to -F(\phi) .
\]
Terms like \eqref{axioncoupling} would generically appear once we
integrate out heavy fermions axially coupled to $\phi$ or as a
result of the Green-Schwarz mechanism \cite{GS}. In the Appendix we will
explicitly show how  a linear $F$ of the form
\be\label{Fval}
       F(\phi) =\frac{\cal N}{(16\pi^2 M_\Pl)}\phi  ,
\ee
 with ${\cal N}$ depending on the details of string
 compactification, arises from heterotic string theory compactified
 to four dimensions \cite{A-G}.

\subsection{Out-of-equilibrium}

Out of equilibrium in our model is achieved noting that   we apply
the interaction in \eqref{axioncoupling}
to the dynamics of metric fluctuations during {\it inflation} where due to (exponential)
growth of the background space-time, the lepton number production is naturally out of
equilibrium.

We are now ready to compute the amount of the
lepton number generated during inflation. In our analysis, we assume that we have a
given successful model of inflation which involves an axion field and do not
elaborate on the details of the inflationary model.  The axion driven models of inflation are known as
natural inflation and are preferred from a particle physics perspective.  Natural inflation
gives a small self coupling for the inflaton field without fine tuning \cite{natural}.
Throughout this note we use notations and conventions of \cite{LL} and in particular,
we use the reduced Planck mass $M_\Pl = 2.44 \times 10^{18}$ GeV.

\section{Gravity wave evolution}

  To begin, we must compute the production of gravitational waves
during inflation under the influence of the coupling \eqref{axioncoupling}.
The action which describes the gravity waves is hence
\be\label{gravity-action}
{\cal L}= \frac{1}{2}M^2_{pl} \sqrt{-\det\ g}\ R+ F(\phi)
R\tilde{R}.
\ee

In general  metric perturbations about an FRW universe can be
parameterized as
\be
\begin{split}
ds^2 &= -(1+2\varphi)dt^2+w_i dtdx^i\cr
&+a^2(t)\left[\left((1+2\psi)\delta_{ij}+h_{ij}\right)dx^idx^j\right]
\end{split}
\ee
where $\varphi$, $\psi$, $w_i$ and $h_{ij}$ respectively parameterize the
 scalar, vector, and tensor fluctuations of the metric.  It is
straightforward to
show that the scalar and vector perturbations do not contribute to
$R\tilde R$, and so we ignore these fluctuations in the following discussion.
We can also fix a gauge so that the tensor fluctuations are parameterized by the
two physical transverse traceless elements of $h_{ij}$.  For such physical gravity waves
which are moving in the $z$ direction, the metric takes the form
\bea\label{mytensors}
ds^2&=&-dt^2+a^2(t)\bigl[(1-\hp)dx^2\cr
&+&(1+\hp)dy^2 + 2\hc dxdy +dz^2\bigr]
\eea
where $a(t)  = e^{Ht}$ during inflation and $\hp$, $\hc$ are functions of
$t$, $z$.

To see the CP violation more explicitly, it is convenient to use a
helicity basis
\be
h_L = \frac{1}{\sqrt{2}} (\hp - i \hc) \ , \quad
h_R = \frac{1}{\sqrt{2}} (\hp + i \hc) \ .
\ee
Here $h_L$ and $h_R$ are complex conjugate scalar fields. To be very explicit,
the negative frequency part of $h_L$ is the conjugate of the positive
frequency part of $h_R$, and both are built from wavefunctions for left-handed
gravitons.
\subsection{The equations of motion}

Plugging \eqref{mytensors} into the action \eqref{gravity-action}, up to second order
in $h_L$ and $h_R$, we obtain
\be\label{RRdual}
\begin{split}
{\cal L}&= -({h_L} \Box {h_R} + {h_R} \Box {h_L})\cr
& +{16i F(\phi)}\biggl[\left(\frac{\partial^2}{\partial z^2}{h_R}
\frac{\partial^2}{\partial t\partial z}{h_L} -
\frac{\partial^2}{\partial z^2}{h_L}
\frac{\partial^2}{\partial t\partial z}{h_R} \right)\cr
& + a^2 \left(\frac{\partial^2}{\partial t^2}{h_R}
\frac{\partial^2}{\partial t\partial z}{h_L} -
\frac{\partial^2}{\partial t^2}{h_L}
\frac{\partial^2}{\partial t\partial z}{h_R} \right) \cr
&  + Ha^2\left(
\frac{\partial}{\partial t}{h_R}\frac{\partial^2}{\partial t\partial z}{h_L}
-\frac{\partial}{\partial t}{h_L}\frac{\partial^2}{\partial t\partial
z}{h_R}\right)\biggr] +{\cal O}(h^4)
\end{split}
\ee
where
\[
\Box= \frac{\partial^2}{\partial t^2}+3H \frac{\partial}{\partial
t}-\frac{1}{a^2}\frac{\partial^2}{\partial z^2}.
\]

As it is explicitly seen from \eqref{RRdual},
if $h_L$ and $h_R$ have the same {\it dispersion relation}, $R\tilde R$ vanishes.
Thus,  nonzero $R\tilde R$ requires
 ``cosmological birefringence'' during inflation.  Such an effect is induced
by the addition of \eqref{axioncoupling} to the gravitational
equations.  Lue, Wang, and
Kamionkowski (LWK) \cite{LWK} and Alexander and Martin \cite{A-M}
have studied the effects of such an interaction
in generating observable parity-violation in the cosmic microwave
background.  In a future work, this coupling will be related to an observable to detect
birefringence in binary systems for the LISA and Advanced LIGO gravitational wave detectors
\cite{Nico}.

It is straightforward to obtain equations of motion for $h_L$ and $h_R$:
\be\label{LReqs}
  \ybox\, h_L = - 2i \frac{\Theta}{ a} {\dot h}^\prime_L \ , \qquad
  \ybox\, h_R = + 2i \frac{\Theta}{a} {\dot h}^\prime_R \ ,
\ee
where
\be\label{Thetaval}
\begin{split}
  \Theta &= \frac{4}{a^2} \frac{d}{dt}(\dot F a^2)/M_{Pl}^2 \\
         &\simeq 4(F''\dot\phi^2+2F'H\dot\phi)/M^2_{Pl}
\end{split}
\ee
dots denote time derivatives, and primes denote differentiation of $F$
with respect to $\phi$. To obtain the above equations we have used the fact that
the inflaton field is only a function of time $t$.
In the second line of the expression for $\Theta$, assuming the slow-roll inflation,
we have dropped the terms proportional to the $\ddot\phi$ (explicitly
we have dropped  $4F'\ddot \phi/M^2_{Pl}$).

Note that  with a constant $\phi$, \eqref{axioncoupling} is the Gauss-Bonnet term and
being a total divergence (in four dimensions), cannot affect the equations of motion;
thus, all terms in $\Theta$ involve derivatives of $\phi$. These  equations should be
compared to those for evolution in flat space
given by LWK~\cite{LWK}.
The new term proportional to $H\dot\phi$
leads to a substantial
enhancement in the size of $\Theta$.   With this
simplification, and the approximate form \eqref{Fval},
\be\label{myTheta}
\Theta = \frac{\sqrt{2\epsilon}}{2\pi^2} \left(\frac{H}{M_{Pl}}\right)^2{\cal N} \ ,
\ee
 where
$\epsilon = \half (\dot\phi)^2/(H M_\Pl)^2$
is the slow-roll parameter of inflation~\cite{LL}.

\subsection{Gravitational birefringence}

To see gravitational birefringence we need to solve the equations of motion explicitly.
Let us focus on the evolution of $h_L$ and, more specifically,
on its positive frequency component.
It is convenient to introduce conformal time
\be\label{eta-t}
          \eta = {1\over H a}  = {1\over H} e^{-Ht} \ .
\ee
(Note that conformal time $\eta$ runs in the opposite direction from $t$.)
The evolution equation for $h_L$ then becomes
\be\label{hLformula}
  {d^2\over d\eta^2} h_L - 2 {1\over \eta} {d\over d\eta} h_L
- {d^2\over dz^2} h_L
          =  -2i\Theta {d^2\over d\eta dz} h_L\ .
\ee

If we ignore $\Theta$ for the moment and let $h_L \sim e^{ikz}$, this
becomes the equation of a spherical
Bessel function:
\be
  {d^2\over d\eta^2} h_L - 2 {1\over \eta} {d\over d\eta} h_L  + k^2 h_L = 0
\ee
for which the positive frequency solution is
\be\label{Bessel}
h_L^{+}(k,\eta) =  e^{+ i k(\eta+z)} (1 - i k\eta)  \ .
\ee

We now look for solutions  to \eqref{hLformula}
with $h_L \sim e^{ikz}$.  To do this, let
\be\label{gdef}
     h_L = e^{ikz} \cdot (-ik\eta)
                        e^{k\Theta\eta}  g(\eta)
\ee
where  $g(\eta)$ is a Coulomb wave function, i.e.
\be\label{Coulomb}
   {d^2\over d\eta^2} g + \left[ k^2 (1-\Theta^2) - {2\over \eta^2}
                 - { 2 k \Theta\over \eta} \right]\, g\ =\ 0 \ .
\ee
This is the equation of a Schr\"odinger particle with $\ell = 1$ in a
weak Coulomb potential.
When $\Theta = 0$, the Coulomb term vanishes and we
find the spherical Bessel function \eqref{Bessel}.
For $h_L$, the Coulomb term is repulsive; for $h_R$, with the opposite sign
of the $\Theta$ term, the Coulomb potential is attractive.
This leads to attenuation of $h_L$ and amplification of $h_R$ in the early
universe. This is  the anticipated cosmological birefringence which was also discussed
by LWK~\cite{LWK}. (For a more detailed treatment see \cite{A-M}.)

As we will see  generation of the matter asymmetry is dominated
by modes at short distances (sub-horizon modes) and at early times. This
corresponds to the limit $k\eta \gg 1$.
In this region, we can ignore the potential
terms in \eqref{Coulomb} and take the solution to be approximately
a plane wave.  More explicitly,
\be\label{findg}
    g(\eta) =   \exp[ ik(1-\Theta^2)^{1/2} \eta (1 + \alpha(\eta))] \ ,
\ee
where $\alpha(\eta) \sim \log \eta/\eta$.

\section{The Green's function}

We would now like to use \eqref{gdef} to compute
the expectation value of $R\tilde R$ in the inflationary space-time.
It turns out that it will be dominated by the sub-horizon, {\it quantum} part of the
gravity-wave evolution. Hence, to compute $\langle R\tilde R\rangle$ we only need the
two point (Green's) function $\langle h_L h_R\rangle$:
\be\label{Gdefin}
\begin{split}
   G(x,t;x',t') &= \VEV{h_L(x,t) h_R(x',t')} \cr
                & = \int \, {d^3 k\over (2\pi)^3}
                 e^{ik \cdot (x-x')} G_k(\eta,\eta')\ .
\end{split}
\ee
For $k$ parallel to $z$, the Fourier component $G_k$ satisfies
\eqref{hLformula} with a delta-function source
\be\label{Gequation}
  \left[{d^2\over d\eta^2} - 2 ({1\over \eta}+k\Theta) {d\over d\eta}
+ k^2 \right]G_k(\eta,\eta')  =
          i  {(H\eta)^2\over M_\Pl^2} \delta(\eta - \eta') .
\ee
For $\Theta = 0$, the solution of this equation is
\be\label{Gvalue}
   G_{k0}(\eta, \eta') = \left\{\begin{array}{cc}
           \aleph   h_L^{+}(k,\eta) h_R^{-}(-k,\eta')
                   &   \quad \eta < \eta' \cr   \ \ \ &\ \ \\
           \aleph    h_L^{-}(k,\eta) h_R^{+}(-k,\eta')
                   &   \quad \eta' < \eta\ ,
\end{array}\right.
\ee
where
\be
\aleph \equiv \left(\frac{H}{M_{Pl}}\right)^2\frac{1}{2k^3}
\ee
and $h_L^{-}$ is the complex conjugate of \eqref{Bessel}, and
$h_R^{+}$, $h_R^{-}$ are the corresponding solutions of the $h_R$ equation.
For $\Theta = 0$, these solutions are the same as for $h_L$, but the
structure of \eqref{Gvalue} will be preserved when we go to the case
$\Theta \neq 0 $.
The leading effect of $\Theta$ is to introduce
the exponential dependence from \eqref{gdef},
\be\label{myGk}
      G_k =
e^{-k\Theta\eta}  e^{+k\Theta \eta'}  G_{k0}
\ee
for both $\eta >\eta'$ and $\eta <\eta'$.
The prefactor is modified in order $\Theta^2$, and the wavefunctions
acquire additional corrections that are subleading for $k\eta \gg 1$.
Neither of these effects will be important for our result.

The Green's function \eqref{myGk}
can now be used to contract $h_L$ and $h_R$ to
evaluate the quantum expectation value of $R\tilde R$.  The result is
\be\label{RRdualval}
  \VEV{R\tilde R} =  {16\over a^4}\, \int \, {d^3 k\over (2\pi)^3}\
  {H^2\over 2 k^3 M_\Pl^2}
                      \cdot k^4 \Theta  + {\cal O}(\Theta^3)
\ee
where we have picked up only the leading behavior for  $k\eta \gg 1$.

We would like to emphasize  that our expression for $\VEV{R\tilde R}$
is nonzero because of the effect of inflation in producing a CP
asymmetry out of equilibrium.  The
original
quantum state for the inflaton might have had nonzero amplitude for a
range of values of $\phi$ and might even have been CP-invariant.
However, inflation collapses the wavefunction onto a particular value of
$\phi$ that is caught up in the local expansion of the universe.  This
value gives us a classical background that is CP-asymmetric.

The above result and computations seems to be crucially dependent on the form of the
Green's function or the vacuum state we have used. To resolve the possible ambiguity
in this regards, one may  perform
the above computation using a different method, the fermion level crossing, e.g.
following \cite{Gibbons-Steif}. This computation  confirms the above results \cite{to-appear}.

\section{Computing the ratio of lepton to photon densities}

To complete our leptogenesis model, there are two  remaining steps. First,
we need to compute the net lepton anti-lepton asymmetry generated. This, in our model,
should be computed during inflation. We then need to have a reheating model through which
we can compute the temperature and hence the entropy of the Universe after the inflation ended.

\subsection{Lepton number density}

We are now ready to evaluate the lepton density that arises through the gravitational
anomaly \eqref{Jlepton}. Inserting \eqref{RRdualval} into \eqref{Jlepton} and
integrating over the time period of inflation, we obtain
\be\label{netlept}
  n = \int^{H^{-1}}_0 d\eta\ \int \, {d^3 k\over (2\pi)^3}\
       {3\over 16\pi^2}\, {16 H^2  k  \Theta \over M_\Pl^2}  \,,
\ee
where $n$ is the lepton number density.
The integral over $k$ runs over all of momentum space, up to the
scale $\mu$ at which our effective Lagrangian description breaks down; that is the momentum integral range is $1<k\eta<\mu/H$.
The dominant contribution comes not from the modes which have become classical and left the horizon by
the end of inflation (the super-horizon modes), $k/(aH) < 1$, but rather from very short distances compared
to these scales; the main contribution comes from subhorizon ``quantum modes.'' As the integrand also shows, the dominant contribution to the
right-hand side comes from $1\ll k\eta <\mu/H$, and as we had anticipated.

We can now  perform the integrals to find the lepton number density produced by the end of
inflation \footnote{In the original version of this paper, due to a computational mistake, we had $(\mu/H)^6$ rather than $(\mu/H)^4$. We thank Azadeh Maleknejad for pointing out this mistake.}
\be\label{nval}
 n  =  {N\over 48\pi^4} \left({ H\over
            M_\Pl}\right)^2 \Theta H^3  \left({\mu\over H} \right)^4 \ .
\ee Let us now analyze each factor in $n$:\\
$\bullet$ The factor
$(H/M_\Pl)^2$ is the  magnitude of the gravity wave power
spectrum. We should stress that the usual gravity wave power
spectrum is coming from the super-horizon modes, while the main
contribution to $n$ has come from the sub-horizon modes.\\
$\bullet$ The
factor $\Theta$ is  a measure of the effective CP violation caused by birefringent
gravity waves.\\
$\bullet$ The factor $H^3$ is the inverse horizon size at inflation; this gives
the density $n$ appropriate units.\\
$\bullet$ Finally, the factor
$({\mu/H})^4$ gives the enhancement over one's first guess due to
our use of strongly quantum, short distance fluctuations to
generate $R\tilde R$, rather than the super-horizon modes which
effectively behave classically. It is intriguing to note that the vacuum energy of gravity mode also grows like $(\mu/H)^4$. 

\subsection{Photon number density}

To understand the significance of the lepton number density \eqref{nval},
we should compare it to the entropy density of the Universe just after reheating,
or to the photon number density. (Recall that almost all of the entropy of the Universe
is generated during the reheating time and is carried in the
massless degrees of freedom, i.e. photons.) In order this we need a reheating model.

To illustrate the physical relevance of the result of our model,
we take the simplest (and at the same time very naive) reheating
model, the instant reheating model. That is, we assume that all of
the energy of the inflationary phase has been converted to the
heat of a  gas of massless particles instantaneously. The
reheating converts the energy density of the inflaton field $\phi$%
\be\label{energy-density}%
\rho= 3 H^2 M_{Pl}^2 %
\ee%
to radiation with%
\be \rho = \pi^2 g_* T^4/30%
\ee%
and to the entropy density $s$,
$s =  2\pi^2 g_* T^3/45$, where $T$ is the reheating temperature, $g_*$ is the effective
number of massless degrees of freedom and $s=1.8g_*\ n_\gamma$ \cite{LL}.
(Here we have assumed that the evolution of the Universe after the reheating era has been
adiabatic.) This gives
\[
n_\gamma =  1.28 g_*^{-3/4} (H M_{Pl})^{3/2}.\]
Recalling that the ratio of the present baryon number to the lepton number originally generated in
leptogenesis is approximately $n_B/n_L = 4/11$ \cite{KRS}, in our model we obtain
\be\label{novers}
    \frac{n_B}{n_\gamma} = 1.8\times 10^{-4}  g_*^{3/4} \left(\frac{H}{M_{Pl}}\right)^{7/2}
                       \ \Theta   \left(\frac{\mu}{H}\right)^4\,. 
\ee
If we are less naive, we might follow the dilution of $n$ and $\rho$ with
the expansion of the universe to the end of reheating.  The final result is
the same (see, however, \cite{Linde} for a comment on this point).
With the adiabatic expansion assumption, \eqref{novers}
 can be compared directly to
the present value of $n_B/n_\gamma$ given in \eqref{baryond}.

\subsection{Numerical results}

We should now estimate the numerical value of $n_B/n_\gamma$ in our  model
to see if we have a viable leptogenesis model. In our final result \eqref{novers} we have
five dimensionless parameters, $g_*,\ H/M_{Pl},\ \mu/M_{Pl}$ and the slow-roll parameter $\epsilon$ and ${\cal N}$
(the latter two  appear through the CP violation parameter $\Theta$). Inserting the expression for $\Theta$
we have
\be\label{n/n-final}
\frac{n_B}{n_\gamma} \simeq 1.3\times 10^{-5} g_*^{3/4}\sqrt{\epsilon}\ {\cal N} \left(\frac{H}{M_{Pl}}\right)^{3/2}
\left(\frac{\mu}{M_{Pl}}\right)^4  .
\ee

Within the usual supersymmetric particle physics models $g_* \sim 1000$ is a reasonable choice.
The WMAP data, through the density perturbation ratio $\delta\rho/\rho$ (for a single field inflation)
leads to an upper bound on $H/M_{Pl}$ ratio as \cite{WMAPtwo}
\be\label{HMPl}
\frac{H}{M_{Pl}}\lesssim 10^{-4}, \qquad    \epsilon \lesssim 0.01,
\ee
or $H\lesssim 10^{14}\ GeV$.

The factor ${\cal N}$ in \eqref{Fval} is inferred from the string theory compactification and is proportional to
the square of the four dimensional $M_{Pl}$ to ten dimensional (fundamental) Planck mass \cite{A-G} (see also the Appendix).
Within string theory,%
\[
{\cal N}\simeq 10^2-10^{10}
\]%
could be a reasonable range.  Therefore, assuming that the
scale of inflation $H$ saturates its current bound $H/M_{Pl}\simeq 10^{-4}$,
\be\label{theta-estimate}
\Theta\simeq 10^{-8}-10^{0}.
\ee

The physically viable range for the parameter $\mu$ depends on the details of the underlying
particle physics model on which our gravi-leptogenesis is based. For example within the Standard Model +
three heavy right handed neutrinos (and the seesaw mechanism) $\mu$ could be of order of the right handed neutrino mass,
in which case $\mu \lesssim 10^{12} GeV$, which is too low for our model to produce the observed matter-antimatter asymmetry.  If we do not
restrict ourselves to the seesaw mechanism, $\mu$ can be larger. The upper bound on $\mu$ is then only coming
from the range of validity of our effective field theory analysis. For example, we can accommodate $\mu$'s as large as
the string scale where quantum gravity effects become important.  Therefore, depending on the details of the model
\[
10^{12} GeV \lesssim \mu \lesssim 10^{16}-10^{17} GeV
\]
or
\be\label{mu-estimate}
10^{-6}  \lesssim \frac{\mu}{M_{Pl}} \lesssim 10^{-2}-10^{-1}
\ee
is a reasonable range for $\mu$ from the particle physics viewpoint. The observed value of matter-antimatter asymmetry \eqref{baryond}, with the value of other parameters as discussed above implies that
$$
3\times 10^{-3}\lesssim \frac{\mu}{M_{Pl}}\sim {{\cal N}}^{-1/4}\lesssim 3\,,
$$
which is well within the particle physics acceptable range \eqref{mu-estimate}.

In sum, assuming that scale of inflation $H$ saturates
its current bound $H\simeq 10^{14} GeV$, we have
\be\label{n-n-estimate}
    \frac{n_B}{n_\gamma} \simeq 3.2
                       \ \Theta \  \left(\frac{\mu}{M_{Pl}}\right)^4 \ .
\ee
Noting \eqref{theta-estimate} and \eqref{mu-estimate}, it is readily seen that
there is big parameter space for obtaining the observed baryon asymmetry of the Universe within our gravi-leptogenesis
model.

\section{Summary and Outlook}

Here we have constructed a leptogenesis model in which the lepton asymmetry is a result of
gravitational chiral anomaly. Indeed our model is a ``module,'' which uses minimal ingredients and could be fit into a
successful inflation model in which the inflaton field(s) has a pseudo-scalar component.
{}From the particle physics side, it is again like a module and many different particle physics models
could be invoked. Here we did not discuss the details of neither the inflationary nor the underlying
particle physics models. It is of course very important to explicitly fit our module
in viable physical models.

As our construction is a leptogenesis model, we need a mechanism to convert the lepton asymmetry to
baryon asymmetry. With the Standard Model this is usually performed via the
thermally activated electro-weak sphalerons, which in our model should
be active after inflation. This happens for reheat temperature $T_r\gtrsim 1$ TeV. Within our
naive ``instant reheating'' model that translates to $H \gtrsim 10^{-3}$ eV, which is not a serious constraint
on the inflationary models used.

According to recent WMAP observations, the scalar metric perturbations generated during
 inflation have a size that gives density fluctuations with
  $ \delta \rho/\rho \sim 10^{-5}$. On the other hand, noting
  \eqref{baryond}, one has
 \be\label{goal}
         {n_B\over n_\gamma} \sim \left({\delta \rho\over\rho} \right)^2 \ .
 \ee
The above could be a simple numerical accident or there might be some underlying deeper physics.
In any case, our model which invokes gravity waves as the source for the baryon asymmetry
and operates during inflation, could be an interesting framework to uncover the possible
deeper physics behind \eqref{goal}.

We would also like to comment on the  dependence of the lepton asymmetry produced on the cutoff $\mu$. The $(\mu/H)^4$ dependence in \eqref{netlept} is intriguingly the same as the cutoff dependence of gravity mode vacuum energy. 

\vspace*{-.7cm}
\begin{acknowledgments}  \vspace*{-.5cm}
We would like to thank Azadeh Maleknejad who pointed out a computational mistake in our earlier version.
We, and in particular M.M. Sh-J who presented the talk, would like to thank organizers of IPM-LHP06 school and conference.
\end{acknowledgments}
\vspace*{-.5cm}


\begin{appendix}

\section{Realization of the model within string theory \cite{A-G}}
\end{appendix}

To see how this term appears in string theory, consider e.g. the Heterotic SUGRA action:
\[
\begin{split}
S=&M_{10}^8\int d^{10}x \sqrt{\det g_{10}}\bigl( {\cal R}
 + \frac{1}{2} (\partial \Phi)^2\\
+& \frac{1}{12}e^{-\Phi} H_{ABC}^2+ \frac{1}{4}e^{-\Phi/2} (F_{AB})^2\bigr)
\end{split}
\]
where
\[
H_3=dB_2-\frac{1}{4}\left(\Omega_3(A)+\alpha'\Omega_3(\omega)\right)
\]
$(\Omega_3(A)$ and $\Omega_3(\omega)$ are the gauge  and gravitational Chern-Simons
three-forms, respectively. Explicitly,
\[
\Omega_3(\omega)=Tr(\omega\wedge d\omega+\frac{2}{3}\omega\wedge\omega\wedge \omega)
\]
In particular note that
\[
^*(d\Omega_3(\omega))=R\tilde{R}.
\]

Upon compactification to four dimensions, the $H^2$ term leads to
\[
\begin{split}
S&\sim M_{10}^8\int d^6y \int d^4 x \ e^{-\Phi} (dB+\alpha'\Omega_3(\omega))^2\\
  &= M_{10}^8 {\frac{V_6}{g_s}} \int d^4x [(dB)^2+2\alpha' (^*dB)\wedge
\Omega_3(\omega)+\cdots]\\
  &= \frac{M_{pl}^2}{g_s} \int d^4x ((\partial\phi)^2 -2\alpha' \phi R\tilde{R})
\end{split}
\]
where the pseudoscalar $\phi$ (our inflaton field) is dual to the two form $B_{2}$ and
the 4d Planck length, $M_{pl}$ is defined as
\[
M_{pl}^2= M_{10}^8V_6.
\]
And hence, finally:
\[
{\cal N}=8\pi^2  \frac{M^2_{pl}
\alpha'}{g_s}=8\pi^2\left(\frac{M_{pl}}{M_{10}}\right)^2\frac{1}{\sqrt{g_s}}
\]


\begin{thebibliography}{99}
\bibitem{Linde}
If the reheating occurs slowly to a reheat temperature $T_r\ll T$,
then $n/s$ may be diluted with respect to this estimate by $T_r/T$. We
thank A. Linde for this comment.
\end{thebibliography}
\end{document}